  \documentclass[preprint2]{aastex}

\newcommand{\mathsym}[1]{{}}
\newcommand{\unicode}[1]{{}}
 \usepackage{graphics}
 \usepackage{graphicx}
 \usepackage{epsfig}

\usepackage{amssymb}
\usepackage{amsmath}
\usepackage{aas_macros}

\shorttitle{Analysis of Equilibrium Points\dots Four Body Problem}
\shortauthors{Reena Kumari \and Badam Singh Kushvah}

\begin{document}


\title{Equilibrium Points and Zero  Velocity Surfaces  in the Restricted Four Body Problem with Solar Wind Drag}


\author{Reena Kumari \and Badam Singh Kushvah}
\affil{Department of   Applied  Mathematics,\\
Indian School of Mines, Dhanbad - 826004, Jharkhand,India}
%
\email{reena.ism@gmail.com; bskush@gmail.com} 

\begin{abstract}
We have analyzed the motion of an infinitesimal mass in the  restricted four body problem with solar wind drag. It is assumed that forces which govern the motion are mutual gravitational attractions of the primaries, radiation pressure force and solar wind drag.  We have derived the equations of motion and find the Jacobi integral, zero velocity surfaces and particular solutions of the system. It is found that three collinear points are real when radiation factor  $0<\beta<0.1$ whereas only one real point obtained when  $0.125<\beta<0.2$. Again, stability property of the system is examined with the help of  Poincar\'{e} surface of section (PSS) and Lyapunov characteristic exponents (LCEs). It is found that in presence of drag forces LCE is negative for a specific initial condition, hence the corresponding trajectory is regular whereas regular islands in the PSS are expanded.
\end{abstract}

\keywords{Restricted four body Problem; radiation pressure; zero velocity surface; P-R drag;solar wind;PSS;LCE}

\section{Introduction}
\label{sec:Int}
In space dynamics there are a number of systems like two body, three body, four body, N-body problem etc. The simplicity and elusiveness of the three body problem in different form like restricted three body problem (RTBP), restricted four body problem (RFBP)(may be consider as an approximation of two three body problem) etc. have attracted the attention of researchers for centuries. The motion of a spacecraft or satellite in the Sun-Earth-Moon system is a simple example of RFBP in space.  The restricted four body has many possible uses in the dynamical system for example, the fourth body is very useful for saving fuel and time in the trajectory transfers in the restricted four body problem \citep{Machuy2007AdSpR..40..118M}.  

The description of the effect of radiation pressure force was first time given by \cite{Poynting1903MNRAS..64A...1P}  and the effect of total radiation force on a particle $P$ due to radiation source $\bf S$  was analyzed  by \cite{Robertson1937MNRAS..97..423R} with the help of  the general relativity theory. He stated that if we consider only first order term in $\frac{\vec v}{c}$ then it consists  a justifiable approximation in classical mechanics to yield \citep{Ragos1995A&A...300..579R}
\begin{eqnarray}
 F=F_p\left(\frac{\vec r}{r}-\frac{\vec v .\vec r}{c r}\frac{\vec r}{r}-\frac{\vec v}{c}\right),\label{eq:rad1}
\end{eqnarray}
where 
 $F_p=\frac{3LM}{16 \pi R^2 \rho A c}$  denotes the measure of the radiation pressure force, $\vec r$ is the position vector of $P$  with respect to the Sun, $\vec v$ is the corresponding velocity vector and $c$ is the velocity of light. In expression of $F_p$,   $L$ is the luminosity of the radiating body whereas $M$, $\rho$ and $A$ are the mass, density and cross section of the particle respectively. In equation (\ref{eq:rad1}) on right hand side, first term expresses  the radiation pressure and second term represents the Doppler shift owing to the motion of the particle whereas third term comes on account of the absorption and subsequent re-emission of the radiation. The last two terms are called P-R drag effect.

 Now, due to the solar wind drag force, equation (\ref{eq:rad1}) can be written as \citep{Burns1979Icar...40....1B}
\begin{eqnarray}
 m\vec{\ddot r}=&&\frac{S A Q_{pr}}{c}\left[\left(1-(1+sw)\frac{\vec v .\vec r}{c ~ r}\right)\frac{\vec r}{r}\right.\nonumber\\&&\left.-(1+sw)\frac{\vec v}{c}\right],\label{eq:rad2}
\end{eqnarray}
 where $S$ and $Q_{pr}$ are the solar energy flux density and the radiation pressure coefficient respectively whereas $sw$ represents the ratio of solar wind drag to Poynting-Robertson(P-R) drag \citep{Liou1995Icar..116..186L}. The velocity independent term in (\ref{eq:rad2}) denotes the radiation pressure however velocity dependent term represents the drag force. The solar wind drag arises from the interaction between solar wind ions and the dust particles. The radiation pressure coefficient $Q_{pr}$ depends on the properties of the particle $P$ and the radiation factor $\beta$ is defined by $\beta=\frac{radiation~ pressure~  force}{solar~  gravitation ~ force}$=$\frac{SAQ_{pr}r_s^2}{c G M m}$.

 Few years ago, \cite{Burns1979Icar...40....1B} discussed about the radiation forces on small particles in the solar system and examined the different types of effect of the radiating body. However,  \cite{Schuerman1980ApJ...238..337S} determined the equilibrium points and examined their stability in the presence of radiation pressure and P-R drag forces.  The dynamical effect of general drag force (i.e. gas drag, nebular drag, PR drag etc.)  in the planar circular restricted three body problem was described by \cite{Murray1994Icar..112..465M}. Also, he has examined the stability of Lagrangian equilibrium points using linear approximation. Further, \cite{Liou1995Icar..116..186L} analyzed the effect of radiation pressure, P-R drag and solar wind drag on the motion of dust grains which is trapped in mean motion resonances with the Sun and the Jupiter in the restricted three body problem and found that all dust gains orbits are unstable.
Again, \cite{Kalvouridis2006AIPC..848..637K} discussed the effect of radiation force due to primaries in the restricted four body problem using Radzievskii's model and noticed that there are some variations in the result which are unstable for all values of the parameters assumed by him. Further, \cite{Ishwar2006math......2467I} and \cite{Kushvah2008Ap&SS.315..231K}, studied the restricted three body problem with P-R drag and examined the effect of P-R drag force on the motion of infinitesimal body. 

Lyapunov characteristic exponent (LCE) and Poincar\'{e} surfaces of section (PSS) are efficient  method to study the behavior of orbit around the equilibrium points. The LCE are very useful tool for the estimation of chaoticity of the trajectory in a  dynamical system. Basically, it measure the rate of exponential divergence  between neighborhood trajectories in the phase space. The basic concepts of LCE was given by \cite{Oseledec1968} during the study of ergodic theory of dynamical system. The description of numerical algorithm to calculate LCE was presented by \cite{Benettin1980Mecc...15....9B} and \cite{Froeschle1984CeMec..34...95F}.
Whereas the determination of the stability regions of the infinitesimal body was first time  introduced by \cite{Poincare1892QB351.P75......} during the study of periodic orbit of the system. This is a good technique to study the nature of trajectory of infinitesimal body and also known as surface of section method. Further, this method was used by \cite{Winter2000P&SS...48...23W} to describe the location and stability of orbit in the restricted three body problem.

 A number of various type of four body models have been studied by many authors \citep{Huang1968VA.....10..113H,Hadjidemetriou1980CeMec..21...63H,Michalodimitrakis1981Ap&SS..75..289M,Rosaev2011arXiv1111.3755R,Ceccaroni2012CeMDA.112..191C} etc. In these models, they have assumed either one or more bodies as radiating and minor body as   very small in comparison to massive bodies. To preserve conservative character these models have been  studied by ignoring dissipative forces like Poynting-Robertson(P-R) drag and solar wind drag.  However, some of the research work in RFBP in addition with radiation pressure, P-R drag etc. have been performed by many authors \citep{Kalvouridis2006AIPC..848..637K,Papadakis2007P&SS...55.1368P,Machuy2007AdSpR..40..118M,Medvedev2008AstL...34..357M} etc. But, little attention  has been paid on the effect of  solar wind drag such models. Therefore, we have considered a restricted four body problem with radiation pressure and solar wind drag.

In order to know the nature of trajectory of infinitesimal body in the proposed problem, we have determined first order LCE for the maximal evolution time $0\leq t\leq10000$ with the help of algorithm given in \cite{Skokos2010LNP...790...63S}. The PSS of the proposed problem is also obtained with the help of Event Locator Method. We have used  Mathematica$^{\textregistered}$ \citep{wolfram2003mathematica} for numerical and algebraic computation.  

The formulation of the problem and derivation of equations of motion are presented in section (\ref{sec:ceqn}) whereas in section (\ref{sec:zvc}), we have obtained the zero velocity surfaces with different cases and location of the collinear equilibrium points is computed in section (\ref{sec:seqn}). Section (\ref{sec:neqn}), we have discussed about the non-collinear equilibrium points however in section (\ref{sec:lce}) we examine the stability behavior of the trajectory in the present dynamical system. Finally, we conclude the paper in section  (\ref{sec:con}).

\section{Formulation of the problem and equations of motion}
\label{sec:ceqn}
We have considered a restricted four body problem in which $m_i, i=1,2,3$ be the masses of three bodies such that $m_3>m_1>m_2$, and $m$ be the mass of infinitesimal body. It is assumed that $m_3$ be a radiating body with solar wind drag which revolves around common center of mass of $m_1$ and $m_2$. 
The governing forces of the motion of infinitesimal body are gravitational attractions of three massive bodies in addition with radiation pressure force, P-R drag and solar wind drag. It is also assumed that the effect of infinitesimal mass on the motion of the remaining system is negligible.
 
We use the canonical system of units, dividing all the distances by the distance between two primaries and dividing all the masses by the total mass of the two primaries. Under these assumption, the mean motion of primaries is taken as unity.

The masses and distances of the Earth, the Moon and the Sun are described as follows: mass of the Earth$(M_E)$ $=5.98 \times 10^{24} kg$; mass of the Moon$(M_M)$ $=7.35 \times 10^{22} kg$;  mass of the Sun$(M_S)$ $=1.99 \times 10^{30} kg$; distance between the Earth and the Moon is $d_1 =3.844 \times 10^5 km;$ distance between the Sun and the Earth is $d_2 =1.496 \times 10^8 km.$ 
Again, the masses of the Earth, the Moon and the Sun in the canonical system are given as $\mu_E = \frac{M_E}{M_M+M_E} = 0.987871$, $\mu_M = \frac{M_M}{M_M+M_E}=0.012151$  and 
 $\mu_S = \frac{M_S}{M_M+M_E} = 328900.48$ respectively. 

\begin{figure}[h]
 \begin{center}
 \plotone{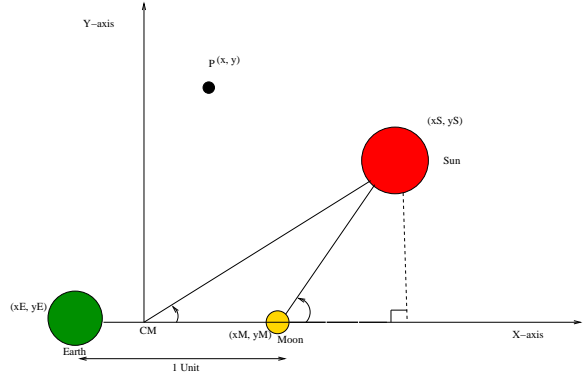}
  \caption{Geometry of the problem} \label{fig:geo}
  \end{center}
  \end{figure}

 Let $\mu_E$ and $\mu_M$ be the radii of the Moon and the Earth respectively and $(x,y),(x_E,y_E),(x_M,y_M)$ and $(x_S,y_S)$ be the co-ordinates of the spacecraft, the Earth, the Moon and the Sun respectively (figure \ref{fig:geo}).

Let $x_E = -\mu_M \cos(t),  y_E=-\mu_M \sin(t),  x_M = \mu_E \cos(t),  y_M = \mu_E \sin(t),  x_S = R_S \cos(\psi),  y_S = R_S \sin(\psi) ~~ \text {and} ~~ \psi = \psi_0+\omega_s t$, where $R_S = 389.1724$ is the distance between the Sun and the center of mass of the system, $\omega_S = 0.07480$ is the angular velocity of the Sun which makes an angle $\psi$ with x-axis, $\psi_0$ is the initial value of $\psi$ and $t$ is the time. The distances of spacecraft from the Earth, the Moon and the Sun are given as  $r_1 = \sqrt{(x-x_E)^2+(y-y_E)^2}$ \quad  $r_2 = \sqrt{(x-x_M)^2+(y-y_M)^2}$  \quad and $r_3 = \sqrt{(x-x_S)^2+(y-y_S)^2}$ respectively.

The equations of motion of the infinitesimal mass in the inertial reference system are
\begin{eqnarray}
 \ddot x& =& -\frac{\mu_E(x-x_E)}{r_1^3}-\frac{\mu_M(x-x_M)}{r_2^3} \nonumber\\&&-\frac{\mu_S (1-\beta)(x-x_E)}{r_3^3}\nonumber\\&&+(1+sw)F_{PR,x},\label{eq:rad3}
\end{eqnarray}
\begin{eqnarray}
 \ddot y &=&-\frac{\mu_E(y-y_E)}{r_1^3}-\frac{\mu_M(y-y_M)}{r_2^3} \nonumber\\&&-\frac{\mu_S(1-\beta)(y-y_E)}{r_3^3}\nonumber\\&&+(1+sw)F_{PR,y},\label{eq:rad4}
\end{eqnarray}where
\begin{eqnarray}
F_{PR,x} =-\frac{\beta \mu_S}{cr_3^2}\left[k_1(x-x_S)+\dot x\right],\\
F_{PR,y} =-\frac{\beta \mu_S}{cr_3^2}\left[k_1(y-y_S)+\dot y\right].\label{eq:rad5}
\end{eqnarray}But,
\begin{eqnarray}
 \nonumber k_1=\frac{\left\{(x-x_S)\dot x +(y-y_S)\dot y\right\}}{r_3^2}
\end{eqnarray}
 In the above expression $q_1=(1-\beta)=(1-\frac{F_p}{F_g})$ and  $c$ are  the mass reduction factor and dimensionless speed of the light respectively whereas $F_p$ and $F_g$  are  the radiation pressure  and  the gravitational force respectively.

Using the coordinate transformation \( x=\xi \cos{t}-\eta \sin{t};\quad y=\xi \sin{t}+\eta \cos{t}\),  equations (\ref{eq:rad3}) and (\ref{eq:rad4}) become in the rotating reference frame as:
 \begin{eqnarray}
\nonumber \ddot \xi-2 \dot \eta = \xi-\frac{\mu_E (\xi-\xi_E)}{r_1^3}-\frac{\mu_M (\xi-\xi_M)}{r_2^3}\\-\frac{\mu_S (1-\beta)(\xi-\xi_S)}{r_3^3}+(1+sw)F_{PR,\xi},\label{eq:33}
\end{eqnarray}
\begin{eqnarray}
\nonumber \ddot \eta+2 \dot \xi = \eta-\frac{\mu_E (\eta-\eta_E)}{r_1^3}-\frac{\mu_M (\eta-\eta_M)}{r_2^3}\\-\frac{\mu_S (1-\beta)(\eta-\eta_S)}{r_3^3}+(1+sw)F_{PR,\eta},\label{eq:rad7}
 \end{eqnarray} or
\begin{eqnarray}
 \ddot \xi-2\dot \eta = \frac{\partial U}{\partial \xi}+(1+sw)F_{PR,\xi},\label{eq:rad8}
\end{eqnarray}
\begin{eqnarray}
\ddot \eta+2\dot \xi = \frac{\partial U}{\partial \eta}+(1+sw)F_{PR,\eta},\label{eq:rad9}
\end{eqnarray}where
\begin{eqnarray}
 U=\frac{\mu_E}{r_1}+\frac{\mu_M}{r_2}+\frac{\mu_S(1-\beta)}{r_3}+\frac{\xi^2+\eta^2}{2},\label{eq:rad10}
\end{eqnarray} 
is the potential in the rotating coordinate system. The P-R drag components in $(\xi,\eta)$  reference system are
\begin{eqnarray}
 \nonumber F_{PR,\xi} = -\frac{\beta \mu_S}{cr_3^2}\left[\left\{\dot \xi-(\eta-\eta_S)\right\}+n_1(\xi-\xi_S)\right] \label{eq:ras1}\end{eqnarray}and
\begin{eqnarray}
\nonumber  F_{PR,\eta} = -\frac{\beta \mu_S}{cr_3^2}\left[\left\{\dot \eta+(\xi-\xi_S)\right\}+n_1(\eta-\eta_S)\right],\label{eq:ras2}
\end{eqnarray}where
\begin{eqnarray}
\nonumber n_1=\frac{\left \{(\xi-\xi_S)\dot \xi+(\eta-\eta_S)\dot \eta \right\}}{r_3^2}.
\end{eqnarray}
If we multiply  the first   equation (\ref{eq:rad8}) by $\dot \xi$ and the second  equation (\ref{eq:rad9}) by  $\dot \eta$ and   adding, we get
\begin{eqnarray}
 \nonumber \dot \xi \ddot \xi+\dot \eta \ddot \eta=\frac{\partial U}{d \xi}\frac{d\xi}{dt}+\frac{\partial U}{d \eta}\frac{d\eta}{dt}+\\(1+sw)\left[\frac{\partial F}{d \xi}\frac{d\xi}{dt}+\frac{\partial F}{d \eta}\frac{d\eta}{dt}\right].\label{eq:ras4}
\end{eqnarray}Since $U$ and $F$ do not depend explicitly on time, the expression on the right hand side is the total time derivative of $U$ and $F$. The left hand side can be expressed in terms of the derivatives of the velocities: 
\begin{eqnarray}
  \frac{1}{2}\frac{d}{dt}\left[\dot \xi^2+\dot \eta^2\right]=\frac{dU}{dt}+(1+sw)\frac{dF}{dt},\label{eq:ras5}
\end{eqnarray}
where
\begin{eqnarray}
 \frac{dF}{dt}=-\frac{\beta \mu_S}{c r_3^2}\left[\frac{\left\{(\xi-\xi_S)\dot \xi+(\eta-\eta_S)\dot \eta\right \}^2}{r_3^2}\right.\nonumber\\ \left.-\left \{(\eta-\eta_S)\dot \xi-(\xi-\xi_S)\dot \eta \right \}+\dot \xi^2+\dot \eta^2\right].\label{eq:ras6}
\end{eqnarray}
Integrating of equation (\ref{eq:ras5}) with respect to time, gives
\begin{eqnarray}
 \nonumber \dot \xi^2+\dot \eta^2&=&2U+2(1+sw)F-C\label{eq:ee1}
\end{eqnarray}or 
\begin{eqnarray}
 C&=&2U+2(1+sw)F-v^2,\label{eq:ee2}
\end{eqnarray}
where $v$ is the velocity of infinitesimal body and $C$ is a constant, called the Jacobi integral. Now, the square of the velocity cannot be negative, the motion of the negligible body is restricted to the region where
\begin{eqnarray}
 \nonumber v^2=2U+2(1+sw)F-C\ge0 \label{eq:ee3}
\end{eqnarray}or
\begin{eqnarray}
 U+(1+sw)F\ge\frac{C}{2}.\label{eq:ee4}
\end{eqnarray}
When the position and velocity of the negligible body are known for some initial moment, then we can obtain the Jacobi integral. Since $U$ and $F$ are functions of position and potential of dissipative force only, condition (\ref{eq:ee4}) tells immediately whether the system can ever reach a given point ($\xi$, $\eta$). This condition does not tell about the shape of the orbit, it only determines the region where the particle could move. Also, condition(\ref{eq:ee4}) shows that the larger value of $C$, the smaller the allowed region.

\begin{figure}[h]
\begin{center}
 \plotone{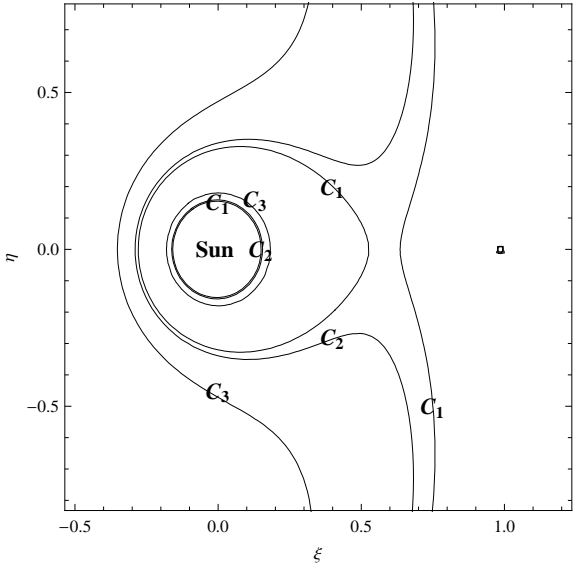}
 \caption{Zero velocity curves at $C_1=848.255,\ C_2=742.489,\ C_3=679.026.$\label{fig:fig10}}
\end{center}
\end{figure}

\begin{figure}[h]
\begin{center}
 \plotone{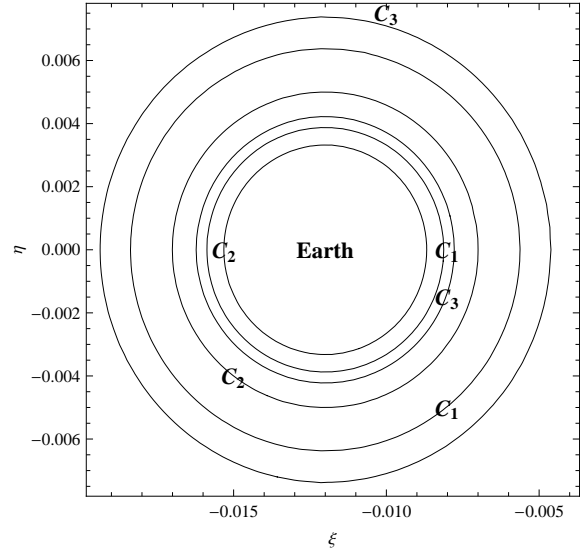}
\caption{Zoom part of figure (\ref{fig:fig10}) at $\xi \in(-0.015, -0.005),\ \eta \in (-0.006, 0.006), \ C_1=848.255,\ C_2=742.489,\ \and C_3=679.026.$ \label{fig:fig12}}
\end{center}
\end{figure}

\begin{figure}[h]
\begin{center}
 \plotone{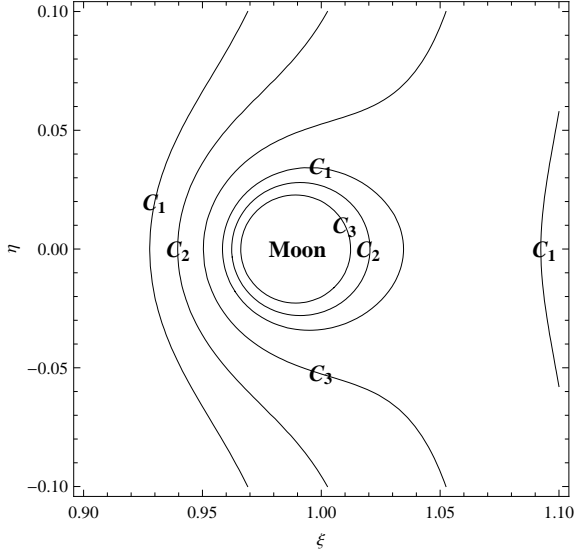}\label{fig:fig8}
\caption{Zoom part of figure (\ref{fig:fig10}) at $\xi \in(0.9, 1.1),\ \eta \in (-0.1, 0.1),\ C_1=848.255,\ C_2=742.489,\ \and  C_3=679.026.$\label{fig:fig11}}
\end{center}
\end{figure}

When $C$ is large, the allowed region consists of the four separate areas. It can never moves form one allowed region to another if the regions are not connected. When $C$ becomes smaller, a connection opens first at the point $L_1$.  Again, we take $C$ even smaller, a connection opens between two inner regions, first at the point $L_2$ and second at the point $L_3$. The small body can never escape from the system. Its orbit is then stable. Finally,  when the value of $C$ is further reduced, the outer and inner regions become connected and escape is possible which can be seen in the figure (\ref{fig:fig10}). It is also observe that how does the connection open with the decreases value of $C$. Figure(\ref{fig:fig12}) and (\ref{fig:fig11}) are the zoom portions of the regions around the Earth and the Moon. The points $L_1$, $L_2$, $L_3$  lie on the same straight line which connects the primaries.


When there are no dissipative forces i.e. $F_{PR,i}=0$, then constant of the motion and  the Jacobi constant $C$ is defined as
\begin{eqnarray}
 C=2U-(\dot \xi^2+\dot \eta^2),\label{eq:ras3}
\end{eqnarray}
which gives the results as in classical model of restricted three body problem.
Again, equation (\ref{eq:ee2}) can be written as
\begin{eqnarray}
 C=2U+b_1+b_2-\dot \xi^2-\dot \eta^2,\label{eq:jc}
\end{eqnarray}where
\begin{eqnarray}
 &&b_1=\frac{2 \beta \mu_S (1+sw)}{c}\left\{\frac{1}{2\left((\xi-389.172)^2+\eta^2\right)}\right.\nonumber\\&& \left.+arctan\left(\frac{\eta}{\xi-389.172}\right)\right\} \quad \text{and} \nonumber\\&& 
b_2=-\frac{2 \beta \mu_S (1+sw)}{c} \int \frac{(\dot \xi^2+\dot \eta^2)}{(\xi-389.172)^2+\eta^2}dt.\nonumber
\end{eqnarray}  
The third term of equation (\ref{eq:jc}) depends on the time due to drag force. Therefore  this integration term shows that Jacobi constant  depends upon the time \citep{murray1999solar, Liou1995Icar..116..186L}.

Consequently, with the help of equation (\ref{eq:ras3}), we obtain the time derivative of Jacobi constant in presence of drag forces which is given as
\begin{eqnarray}
 \dot C&=&2(1+sw)\frac{\beta \mu_S}{c r_3^2}\left\{ \dot \xi^2+\dot \eta^2+\frac{n^2_1}{r^2_3}\right.\nonumber\\&& \left.-\left \{(\eta-\eta_S)\dot \xi-(\xi-\xi_S)\dot \eta\right \} \right\}\label{eq:ras7}
\end{eqnarray}
 We have tried our best to derive the explicitly time independent expression of potential function or Jacobi constant but we could not find the exact analytical expression. In future, this should be taken up as a challenging problem of such models having P-R drag and solar wind drag. 
\section{Zero velocity surfaces}
\label{sec:zvc}
The Jacobi integral is a relation between the square of the velocity and the coordinates of the infinitesimal particle with respect to the set of rotating axes. If the particle's velocity becomes zero, we have
\begin{eqnarray}
 {\xi^2+\eta^2}+\frac{2\mu_E}{r_1}+\frac{2\mu_M}{r_2}+\frac{2(1-\beta)\mu_S}{r_3}
+A_1 = C, \label{eq:as3}
\end{eqnarray}where
\begin{eqnarray}
 A_1=&&\frac{2(1+sw)\beta \mu_S}{c}\left[arctan\left(\frac{\xi-\xi_S}{\eta-\eta_S}\right)\right.\nonumber\\&& \left.-arctan\left(\frac{\eta-\eta_S}{\xi-\xi_S}\right)\right]\end{eqnarray}
 and  $C$ can be determined from the initial conditions.

Equation (\ref{eq:as3}) is important in this problem and it is defined for a given value of $C$ i.e. the boundaries of regions in which the particle is free to move. Now, for large values of $\xi$ and $\eta$, all the terms except first and second in L.H.S. of the equation (\ref{eq:as3}) become unimportant. In other words, this equation takes the form:
\begin{eqnarray}
\xi^2+\eta^2 = C-\sigma = C_1, \label{eq:as5}
\end{eqnarray}where
\begin{eqnarray}
\nonumber \sigma=\frac{2\mu_E}{r_1}+\frac{2\mu_M}{r_2}+\frac{2(1-\beta)\mu_S}{r_3}+A_1.\label{eq:as4}
\end{eqnarray}
The equation (\ref{eq:as5}) represents a circle whose radius is $\sqrt{C_1}$. Therefore the curve in the $\xi \eta$- plane is an approximately oval shape  within the asymptotic cylinder. For small values of $\xi$ and $\eta$,  the first and second terms are relatively unimportant, hence the equation may be written as:
 \begin{eqnarray}
\nonumber \frac{\mu_E}{r_1}+\frac{\mu_M}{r_2}+\frac{(1-\beta)\mu_S}{r_3} +A_1=C_2, \label{eq:as6}
\end{eqnarray}where
\begin{eqnarray}
 C_2=\frac{1}{2}\left(C-\xi^2-\eta^2 \right),\label{eq:as7}
\end{eqnarray}
 which is an equation of equipotential surfaces. 
\begin{figure}[h]
\plotone{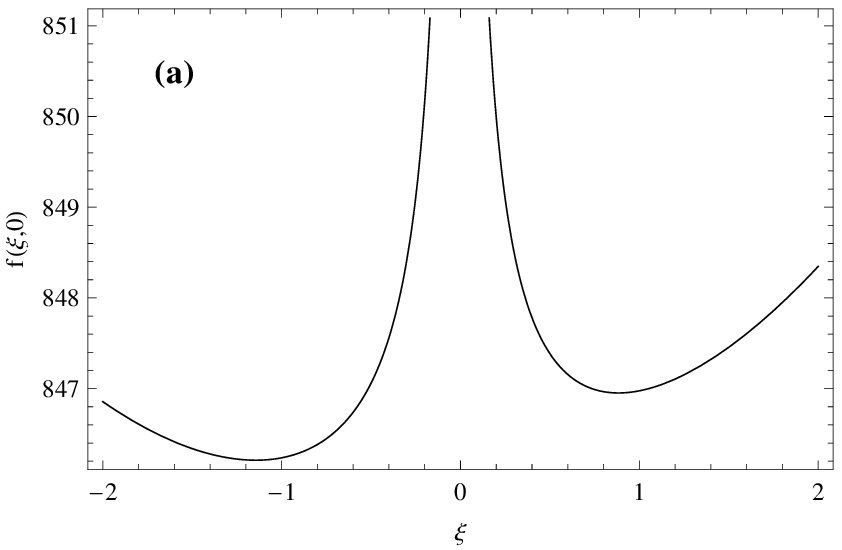}\\ \plotone{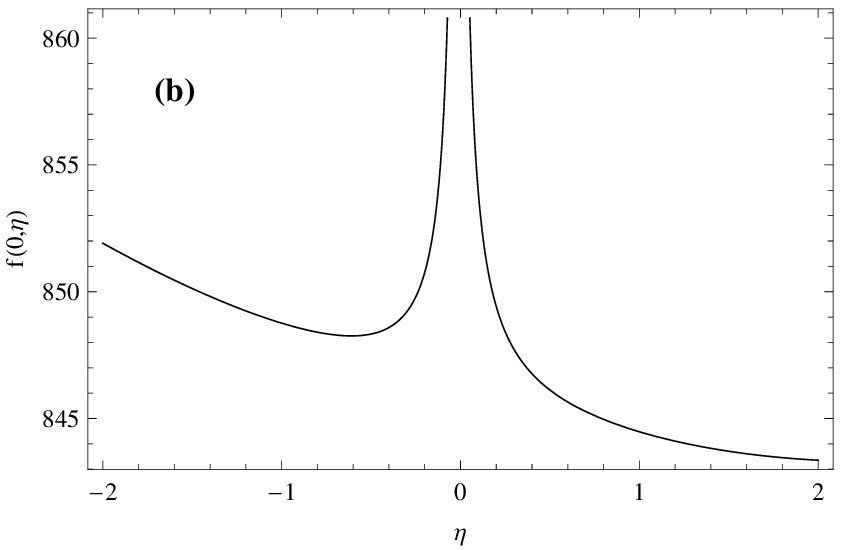}
 \caption{ (a) f($\xi$, 0) and (b) f(0, $\eta$) for  $\mu_E = 0.987,\ \mu_M = 0.012, \ \mu_S = 328900.48, \ t = 1000, \ \psi = 0, \ \beta = 0.00002, \ c = 1.8\times 10^6, \  sw = 0.248.$\label{fig:fig2}}
\end{figure}

\begin{figure}[h]
\plottwo{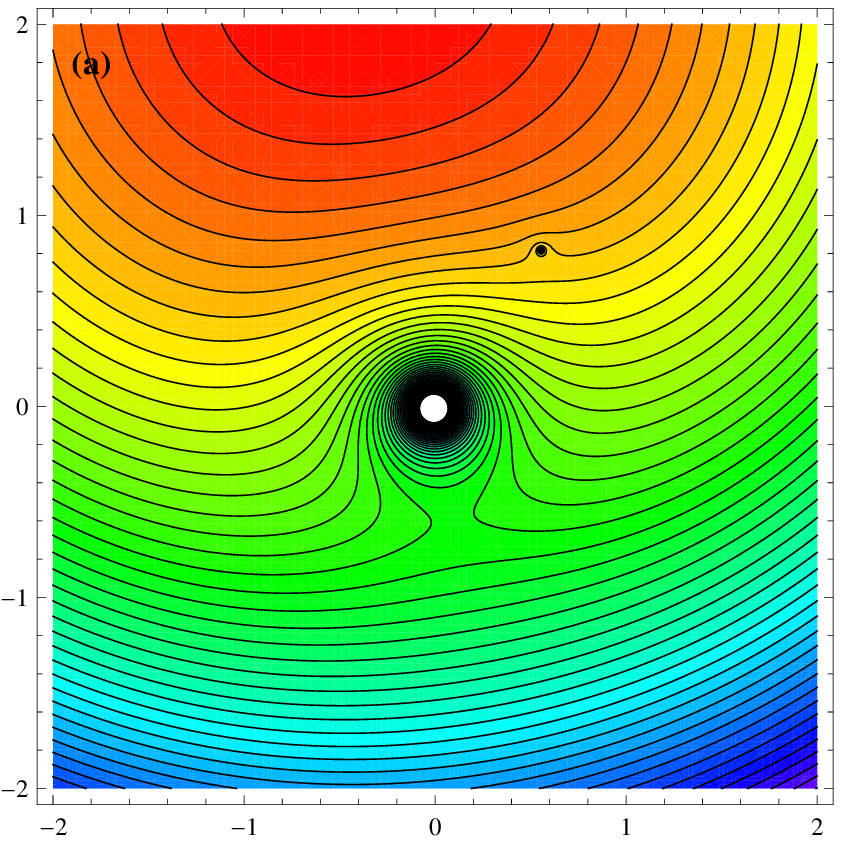}{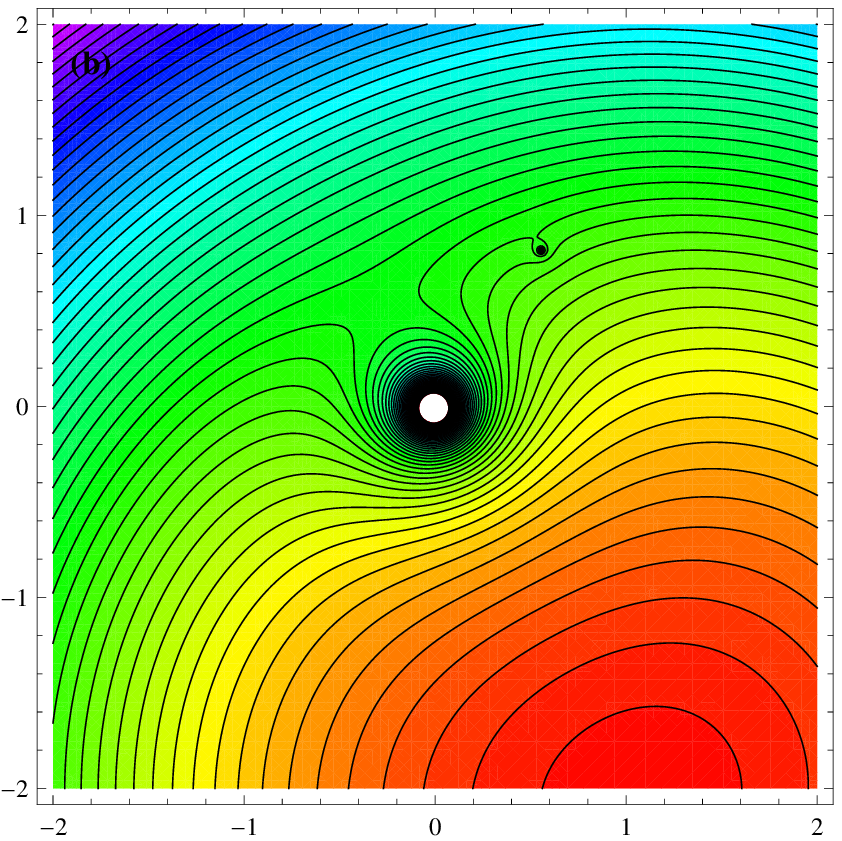} \\ \plottwo{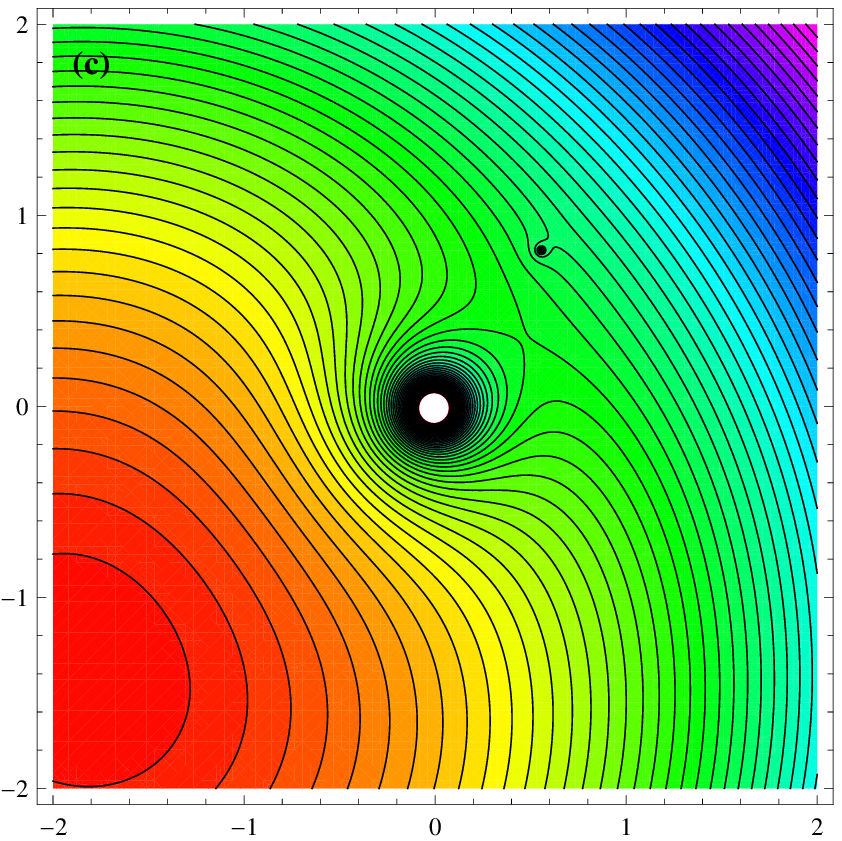}{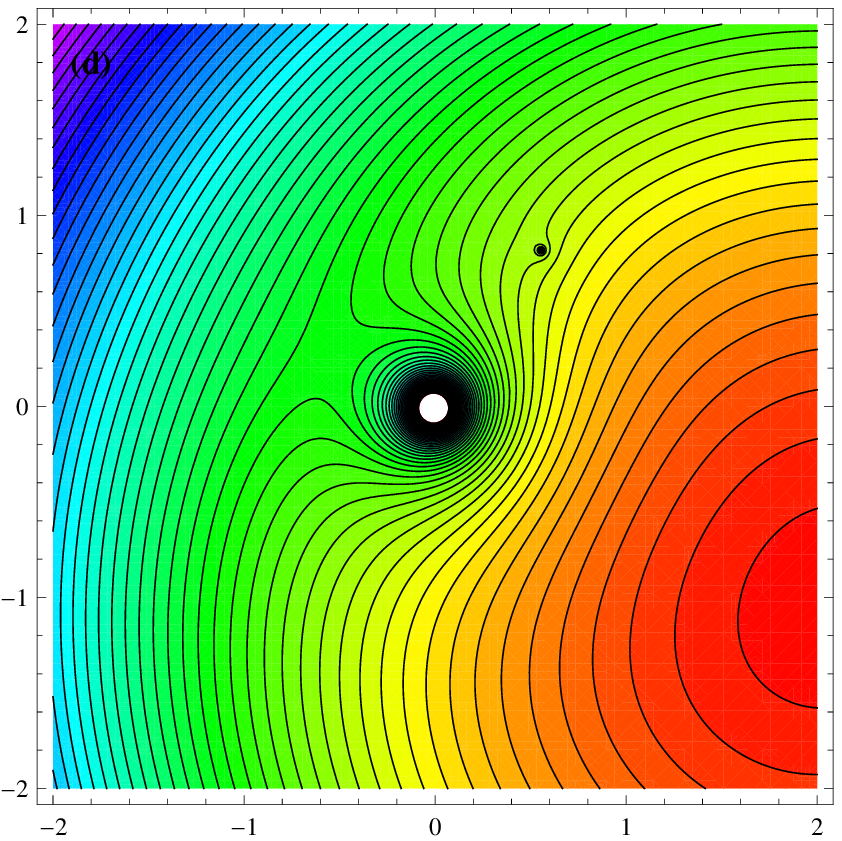}
 \caption{Zero velocity surfaces at (a) $\psi=0$,\  (b) $\psi=60$, \
(c) $\psi=90$ \ and \ (d) $\psi=180$. \label{fig:fig5}}
\end{figure}

Figure (\ref{fig:fig2}), shows that curves intersect at a point where function gives local minimum and  maximum value of Jacobi constant. Again, figure (\ref{fig:fig5}) shows contour curves which are the  equipotential boundaries. It is clear from the frames that the configuration rotates anti-clock wise with an angle $\psi$. In the contour plots red (dark) portion shows the higher equipotential value whereas when we go from red portion to blue (very dark) one, equipotential decreases gradually.
Now, the zero velocity curves are defined by the equation
\begin{eqnarray}
  f(\xi,\eta)=\xi^2+\eta^2+\frac{2 \mu_E}{r_1}+ \frac{2 \mu_M}{r_2}\nonumber \\+\frac{2 (1-\beta)\mu_S}{r_3}+A_1.\label{eq:as8}
\end{eqnarray}
 To determine the curvature of a zero velocity curve(ZVC) (\ref{eq:as8}), we use the formula \citep{Marceta}
\begin{eqnarray}
ZVC= \frac{f_{\xi\xi}f_{\eta}^2-2 f_{\eta\eta}f_\xi f_\eta+f_{\eta\eta}f_\xi^2}{({f_\xi^2}+{f_\eta^2})^{3/2}},\label{eq:as9}
\end{eqnarray}
where
\begin{eqnarray}
 &&f_\xi=\frac{\partial f(\xi,\eta)}{\partial \xi}, f_\eta=\frac{\partial f(\xi,\eta)}{\partial \eta}, f_{\xi\xi}=\frac{\partial^2 f(\xi,\eta)}{\partial \xi^2},\nonumber \\&& f_{\eta\eta}=\frac{\partial^2 f(\xi,\eta)}{\partial \eta^2}, f_{\xi\eta}=\frac{\partial^2 f(\xi,\eta)}{\partial \xi \eta}\nonumber.\end{eqnarray}

Now, we determine the curvature of ZVC (\ref{eq:as8}) with the help of characteristic values of Hessian matrix $H$, i.e. 
\begin{eqnarray}
H=\begin{bmatrix}
    \frac{\partial N_\xi}{\partial \xi}~  \frac{\partial N_\xi}{\partial \eta}\\
\frac{\partial N_\eta}{\partial \xi} ~ \frac{\partial N_\eta}{\partial \eta} \label{eq:aa1}
   \end{bmatrix},
\end{eqnarray}
where $N_\xi$ and $N_\eta$ are components of normal unit vector defined by
\begin{eqnarray}
 \vec N= \left[ \vec N_\xi ~~ \vec N_ \eta \right]=\frac{\vec G}{||G||}\nonumber \\=\frac{\left[\frac{\partial f}{\partial \xi}, \frac{\partial f}{\partial \eta}\right]}{\left[{\left(\frac{\partial f}{\partial \xi}\right)^2+\left(\frac{\partial f}{\partial \eta}\right)^2}\right]^{1/2}},\label{eq:aa2}
\end{eqnarray} with
\begin{eqnarray}
 \vec G=\bigtriangledown f=\left[\frac{\partial f}{\partial \xi} ~ \frac{\partial f}{\partial \eta}\right].\label{eq:aa3}
\end{eqnarray}
 \begin{figure}[h] 
\plotone{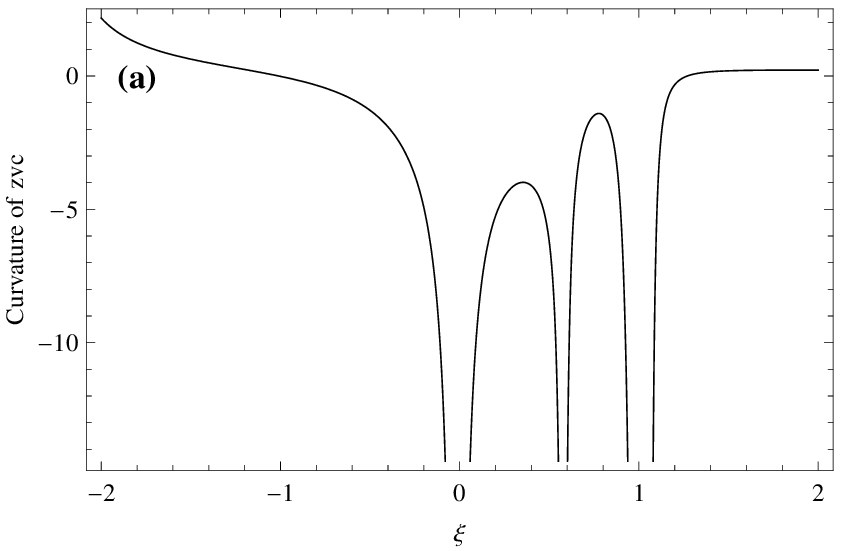}\\
\plotone{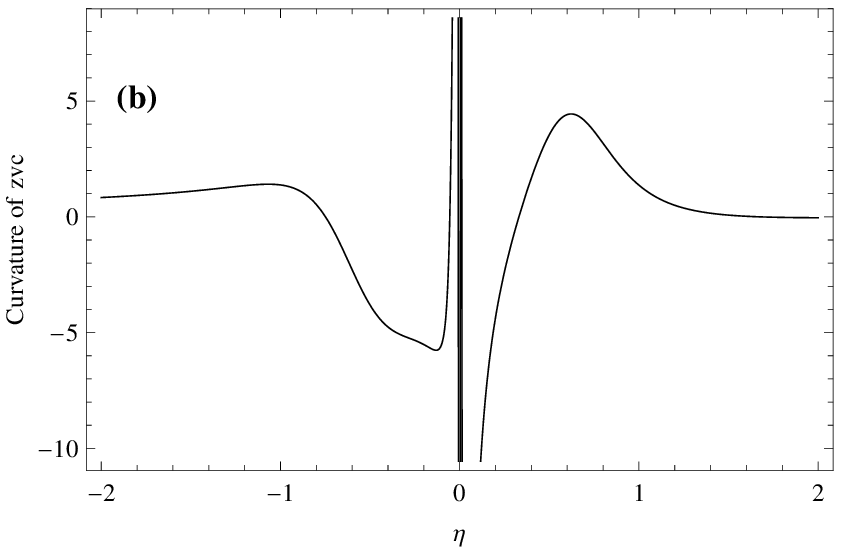}
 \caption{Curvature of (a) f($\xi$, 0) and (b) f(0, $\eta$) for  $\mu_E = 0.987,\ \mu_M = 0.012, \ \mu_S = 328900.48, \ t = 0, \ \psi = 0, \ \beta = 0.00002, \ c = 1.8\times 10^6, \  sw = 0.248.$\label{fig:fig7} }
\end{figure}
Since the expression of the curvature of zero velocity curve is too complicated to deduce anything meaningful analytically. Therefore we have  computed curvatures numerically and  observe  that  the curvature of any curve shows the bending nature   at a particular point(figures  \ref{fig:fig7} ,  \ref{fig:fig8}  and  \ref{fig:fig9}). This value is inverse proportional to the radius of curvature. If curvature is large then radius of curvature is small, that means curve is more bended at that point.
\begin{figure}[h]
\plottwo{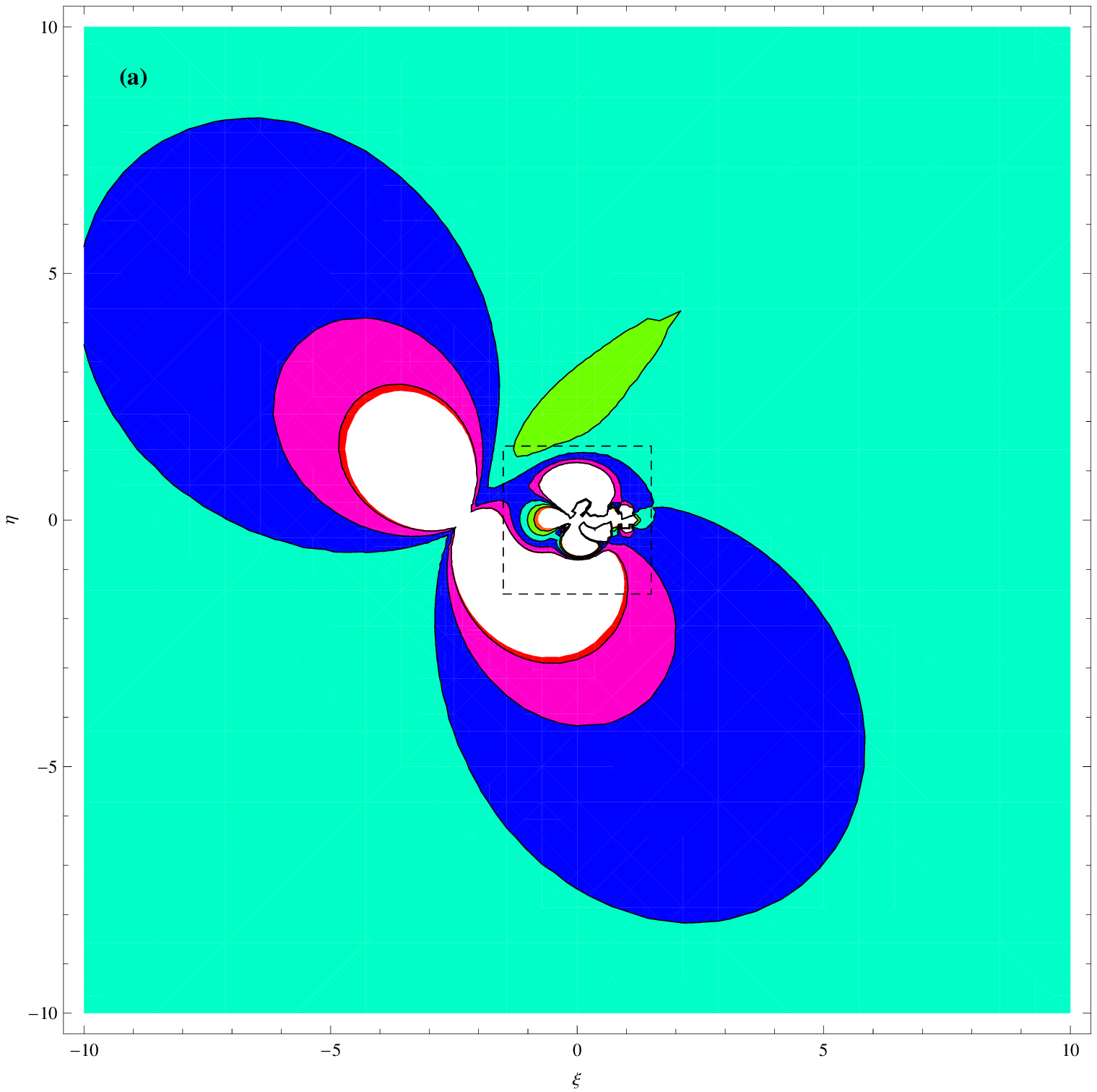}{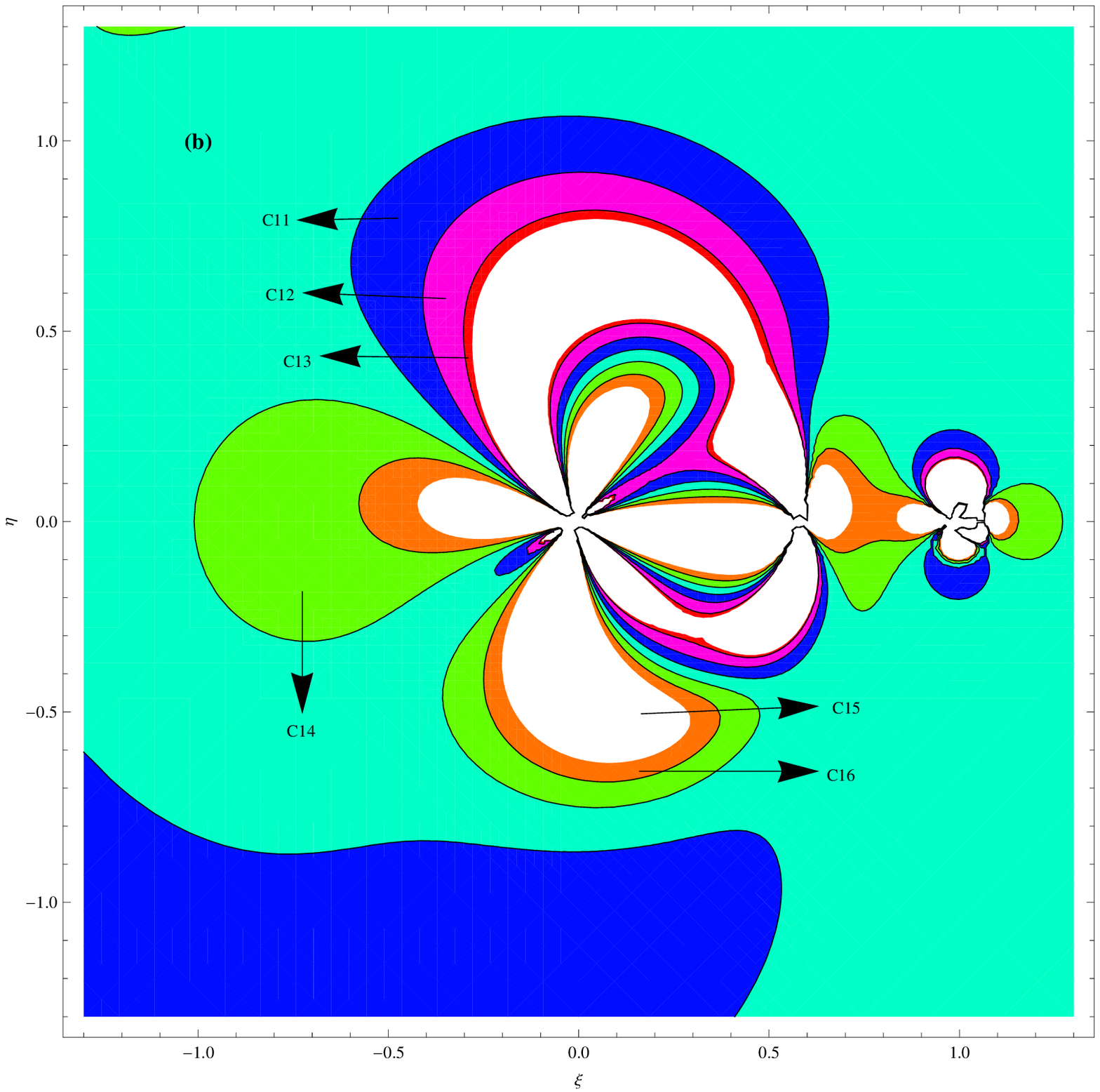}
 \caption{ (a) Curvature of zero velocity surface and (b) Zoom of square portion of figure (a).\label{fig:fig8} }
\end{figure} 
\begin{figure}[h]
\begin{center}
 \plotone{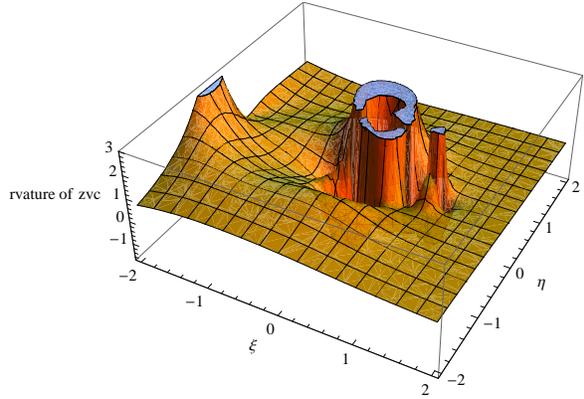}
\end{center}\caption{Curvature of zero velocity surfaces.\label{fig:fig9} }
\end{figure} 
If curvature decreases than radius of curvature increases, that means bending property of the curve decreases which opens the path to move particle freely. From figure(\ref{fig:fig9}) it can be seen that around the singular points, curvature of zero velocity curve is highest that means radius of curvature is least therefore curves are disjoint closed loops form where particles can move form one regions to another.  
In figure (\ref{fig:fig7}), frames (a) and (b) show the curvature of ZVC depends on  $\xi$ and $\eta$ respectively and indicates that the curves are sometimes nearer to each other and sometimes goes far form to each other. When the curves bend to each other then the connection opens where the body moves from one allowed side to another side. When the curve is not bended that means curve goes outside,  then the curves are separate (figure \ref{fig:fig8}). Different layers show the equipotential curvatures where the movement is possible which is clearly seen in zoom part of the figure (\ref{fig:fig8}). Also, figure (\ref{fig:fig9}) shows the curvature of zero velocity surfaces, cut off  portions indicate the singularities of the surfaces where motion of infinitesimal body is impossible.  
  
\section{Location of the Lagrangian equilibrium points}
\label{sec:seqn}
For equilibrium points, we solve the equations (\ref{eq:rad8}) and (\ref{eq:rad9}) by taking $\ddot \xi=0=\ddot \eta=\dot \xi=\dot \eta$ i.e.
\begin{eqnarray}
  \xi-\frac{\mu_E (\xi-\xi_E)}{r_1^3}-\frac{\mu_M (\xi-\xi_M)}{r_2^3}\nonumber \\-\frac{\mu_S (1-\beta)(\xi-\xi_S)}{r_3^3}+(1+sw)F_{PR,\xi} = 0,\label{eq:ras8}
\end{eqnarray}
\begin{eqnarray}
\eta-\frac{\mu_E (\eta-\eta_E)}{r_1^3}-\frac{\mu_M (\eta-\eta_M)}{r_2^3}\nonumber \\-\frac{\mu_S (1-\beta)(\eta-\eta_S)}{r_3^3}+(1+sw)F_{PR,\eta} = 0,\label{eq:ras9}
 \end{eqnarray} where

$F_{PR,\xi}=\frac{\beta \mu_S}{cr_3^2}(\eta-\eta_S), \quad
 F_{PR,\eta}=-\frac{\beta \mu_S}{cr_3^2}(\xi-\xi_S).$

\subsection{Particular cases}
\label{sec:peqn}
1. When $ \beta=0$ then the equation (\ref{eq:ras8}) and (\ref{eq:ras9}) give the following
\begin{eqnarray}                                                 
\nonumber \xi-\frac{\mu_E (\xi-\xi_E)}{r_1^3}-\frac{\mu_M (\xi-\xi_M)}{r_2^3}-\frac{\mu_S (\xi-\xi_S)}{r_3^3} = 0,\\
\eta-\frac{\mu_E (\eta-\eta_E)}{r_1^3}-\frac{\mu_M (\eta-\eta_M)}{r_2^3}-\frac{\mu_S (\eta-\eta_S)}{r_3^3} = 0.
\end{eqnarray}
 This is the classical case of restricted four body problem.

2. If $ \beta=1$ then the equation (\ref{eq:ras8}) and (\ref{eq:ras9}) takes the form 
\begin{eqnarray}
\nonumber \xi-\frac{\mu_E (\xi-\xi_E)}{r_1^3}-\frac{\mu_M (\xi-\xi_M)}{r_2^3}+(1+sw)F_{PR,\xi} = 0,\\
\eta-\frac{\mu_E (\eta-\eta_E)}{r_1^3}-\frac{\mu_M (\eta-\eta_M)}{r_2^3}+(1+sw)F_{PR,\eta} = 0.\end{eqnarray}
This becomes the restricted three body problem with solar wind and P-R drag.
\subsection{Collinear points}
The collinear points are solutions of the equations (\ref{eq:ras8}-\ref{eq:ras9}) with $\eta=0$. That is, the collinear points lie on the line joining the primaries. Therefore, we have
\begin{eqnarray}
 \xi-\frac{\mu_E (\xi-\xi_E)}{r_1^3}-\frac{\mu_M (\xi-\xi_M)}{r_2^3}-\frac{\mu_S (\xi-\xi_S)}{r_3^3} = 0.\label{eq:as1}
\end{eqnarray}
When the value of $\beta$ is zero, the above equation becomes
\begin{eqnarray}
  \xi^7-780.294 \xi^6+152974 \xi^5\nonumber \\-624960 \xi^4+782507 \xi^3-454193 \xi^2\nonumber \\+288215 \xi -145671= 0,\label{eq:as2}
\end{eqnarray}
which shows that the roots lie in the $\xi$ axis and after simplification, the above equation consists seven degree. Using the Descartes' rule of sign we found that three roots are real and other pair of roots are imaginary, because the coefficients of the power of $\xi$ change seven signs. This shows that at a time we can have only three real roots joining the two primaries because the plane of motion of the Earth around the Sun is different from the plane of motion of the Moon around the Earth. The three primaries will be in the same plane when the Moon comes at the line of node of the plane of motion.

Now, if we take $\beta$ non-zero then we obtain three real roots of equation (\ref{eq:as1}) change with the value of  $\beta \in (0, 0.1)$. Again, if $\beta \in (0.125, 0.2)$ then we have only one root which increases with $\beta$. With the help of these values we determine Jacobi constant and see that an increases in the value of $\beta$ consequently the Jacobi constant decreases and the corresponding energy increases (Table \ref{tab:T1}).

\begin{table}
\tiny
\caption{Collinear points for initial parameters $ \mu_E = 0.987$,  $\mu_M = 0.012$,  $\mu_S = 328900.48$,  $t = 0$,  $\psi=0$ \label{tab:T1}}
\begin{tabular}{|rrrrrrr|}
\hline
\bf{$\beta $} & {I } & {II} & {III} & {$C_1$} & {$C_2$} & {$C_3$}  \\\hline

$0.0$&-1.86495&-0.892772&0.578692 &843.374 &844.720 &848.255\\
$0.0000025$  &-1.86494&-0.892775&0.578693& 843.372&844.718 &848.253\\
$0.000005 $  &-1.86494&-0.892778&0.578693& 843.370&844.716 &848.251\\
$0.0000075$  &-1.86493&-0.892780&0.578694& 843.368&844.714 &848.249\\
$0.00001$  &-1.86492&-0.892783&0.578694& 843.366&844.712 &848.247\\
$0.0000125$  &-1.86491&-0.892786&0.578695& 843.364&844.709 &848.245\\
$0.0000150$  &-1.86490&-0.892789&0.578695& 843.362&844.707 &848.242\\
$0.0000175$  &-1.86490&-0.892792&0.578696& 843.359&844.705 &848.240\\
$0.00002$  &-1.86489&-0.892795&0.578696& 843.357& 844.703&848.238\\

$0.025$  &-1.78593&-0.924049&0.583721& 822..399&823.571 &827.103\\
$0.05$  &-1.70064&-0.961693&0.588858& 801.429&802.416 &805.950\\
$0.075$  &-1.60522&-1.009590&0.594104& 780.467&781.248 &784.796\\
$0.10$  &-1.48925&-1.078140&0.599462& 759.527&760.056 &763.643\\
$0.125$  &Complex&Complex&0.604934& Complex&Complex &742.489\\
$0.150$  &Complex&Complex&0.610524&Complex &Complex & 721.335\\
$0.175$  &Complex&Complex&0.616232&Complex &Complex &700.181\\
$0.20$  &Complex&Complex&0.622061& Complex&Complex &679.026\\\hline
\end{tabular}
\end{table}
\subsection{Extremal values}
From equation (\ref{eq:as3}), we have
\begin{eqnarray}
C={\xi^2+\eta^2}+\frac{2\mu_E}{r_1}+\frac{2\mu_M}{r_2}+\frac{2(1-\beta)\mu_S}{r_3}
\nonumber\\+\frac{2(1+sw)\beta \mu_S}{c }\left[\int \frac{(\eta-\eta_S)}{r_3^2}d \xi-\int \frac{(\xi-\xi_S)}{r_3^2}d \eta \right]\label{eq:aa4}
\end{eqnarray}
Now,
\begin{eqnarray}
 &&\mu_E r_1^2+\mu_M r_2^2=\mu_E\left[(\xi-\xi_E)^2+(\eta-\eta_E)^2\right]\nonumber\\&&+\mu_M\left[(\xi-\xi_M)^2+(\eta-\eta_M)^2\right],\nonumber\\&&
\Rightarrow \mu_E r_1^2+\mu_M r_2^2 =\xi^2+\eta^2+\mu_E \mu_M,\nonumber\\&&
\Rightarrow \xi^2+\eta^2= \mu_E r_1^2+\mu_M r_2^2-\mu_E \mu_M.\nonumber
\end{eqnarray}
Substituting these values in equation (\ref{eq:aa4}), we get
\begin{eqnarray}
 &&\phi=\mu_E\left[r_1^2+\frac{2}{r_1}\right]+\mu_M\left[r_2^2+\frac{2}{r_2}\right]\nonumber\\&&+\frac{2(1+sw)\beta \mu_S}{c}\left[\frac{\int (\eta-\eta_S)d \xi}{r_3^2}-\frac{\int (\xi-\xi_S)d \eta}{r_3^2}\right]\nonumber\\&&-\mu_E\mu_M+\frac{2(1-\beta)\mu_S}{r_3}.\label{eq:aa6}
\end{eqnarray}
Differentiating equation (\ref{eq:aa6}) with respect to $r_1, r_2$ and $r_3$  respectively and equating each term to zero for initial values of $t$, we get
\begin{eqnarray}
&& r_1=1,~ r_2=1,\nonumber \\&& r_3=\frac{\beta}{\beta-1}\frac{2 \times (1+sw)\times R_s\times \eta )}{c}.\label{eq:aa7}
\end{eqnarray}
In equation (\ref{eq:aa7}), if we take $r_3=0$, this causes  the Sun to be   placed at the origin which is not possible in the proposed system. Because for a moment if we consider that the Sun is placed at the common center of the Earth and the Moon and consequently, we get a system in which the Earth and the Moon revolve in different orbits about the Sun. In general such type of system does not exist  for the problem in hand.
If $r_3>0$~~ then the factor $\frac{(1+sw)\beta}{(\beta-1)}>0$  has the following cases: 

1. If  $\frac{\beta}{(\beta-1)}(1+sw)$  have positive signs then
$sw> -1$  and $\beta $ lies in the open interval (0, 1) that is $\frac{F_p}{F_g}< 1$.

2. If  $\frac{\beta}{(\beta-1)}$ and $(1+sw)$ have negative signs then
$sw < -1$  and $\beta $ lies outside the open interval (0,1) and the body escapes from the system.

The function $\phi$ always positive for any $r_i, i=1,2,3 $ and $\phi$  approaches to infinity as  $r_i$ approaches to zero or infinity. Therefore, the equilibrium points of the system are absolute minimum however the pseudo-potential and the Jacobi constant at these points are:
\begin{eqnarray}
 \phi=\frac{1}{(1+sw)~\beta~\eta } A_2,\label{eq:aa8}\end{eqnarray}
\begin{eqnarray}
C=\frac{2}{(1+sw) ~\beta~ \eta} A_2,\label{eq:aa9}
\end{eqnarray}where \begin{eqnarray}
 A_2=-7.60616 \times 10^8 
+\beta~\eta \left \{1.52123 \times 10^{9}
 \right.\nonumber \\ \left.+(2.98516+2.98516 \times sw) \right\}. \end{eqnarray}

\subsection{Non-collinear points}
\label{sec:neqn}
For non-collinear points, we take perturbation in distances  of infinitesimal mass from the primaries due to the radiation pressure and solar wind drag and suppose   $r_1 = 1+\epsilon_1$ and $r_2 = 1+\epsilon_2$, where $\epsilon_1,\epsilon_2\ll 1$,   then from center of mass property, we obtain
\begin{equation}
      \mu_Er_1+\mu_Mr_2 = 0\Rightarrow \mu_E\epsilon_1+\mu_M\epsilon_2+1 = 0.\label{eq:aw1}                                                                                                                                                                                                                                                                                    
   \end{equation}
 Now, \( r_1^2-r_2^2 = (\xi+\mu_M)^2+\eta^2-(\xi+\mu_M-1)^2-\eta^2\), therefore
\begin{eqnarray}
 \xi = \frac{2(\epsilon_1-\epsilon_2)+1}{2}-\mu_M.\label{eq:aw2}
\end{eqnarray}
Substituting this value in the equation (\ref{eq:ras8}), we get  
\begin{eqnarray}
\eta= \pm\sqrt{\frac{-(1-3sw-4\beta)\times  c^2\times (\xi-389.172)}{2 \times (1-sw-2\beta)\times 389.172\times \beta^2}}.\label{eq:ss}
\end{eqnarray}
 Again, substituting the value of $\xi$ in the equation (\ref{eq:ras8}) and  neglecting  second and higher order terms of $\beta$, we get

\begin{eqnarray}
 \mu_S (1-4\beta)(1-3sw)c^3\left(\frac{2(\epsilon_1-\epsilon_2)+1}{2}\right.\nonumber\\ \left.-\mu_M-389.172 \right) = 0. \label{eq:aw4}
\end{eqnarray}
Solving equation (\ref{eq:aw1}) and (\ref{eq:aw4}), we get the values of  $\epsilon_1$ and $\epsilon_2$ which are given below
\begin{eqnarray}
 \nonumber \epsilon_1=\frac{1}{-1+3 sw+4\beta} A_3,\label{eq:aw5}
\end{eqnarray}
 \begin{eqnarray}
  \epsilon_2=\frac{1}{-1+3 sw+4\beta} A_4, \label{eq:aw6}
\end{eqnarray}where
\begin{eqnarray}
A_3=-12(0.0833-0.3333 \beta-0.25 sw)\times \nonumber \\(0.0026+\mu_M)\times (388.669+\mu_M)\end{eqnarray}and
\begin{eqnarray}
 A_4=1-3 sw-4\beta\nonumber+(4664.06+12 \mu_M)\times \nonumber &&\\ \mu_E (0.0833-0.3333\beta-0.25 sw).
\end{eqnarray}

Substituting the values of $\epsilon_1$ and $\epsilon_2$  in equation (\ref{eq:aw2})  and (\ref{eq:ss}), we get the non-collinear points. These points depend on the values of the radiation pressure and the solar wind drag. When we change the values of ratio of solar wind to P-R drag parameter for different fixed values of radiation pressure parameter, points are changes.
\begin{figure}
 \plotone{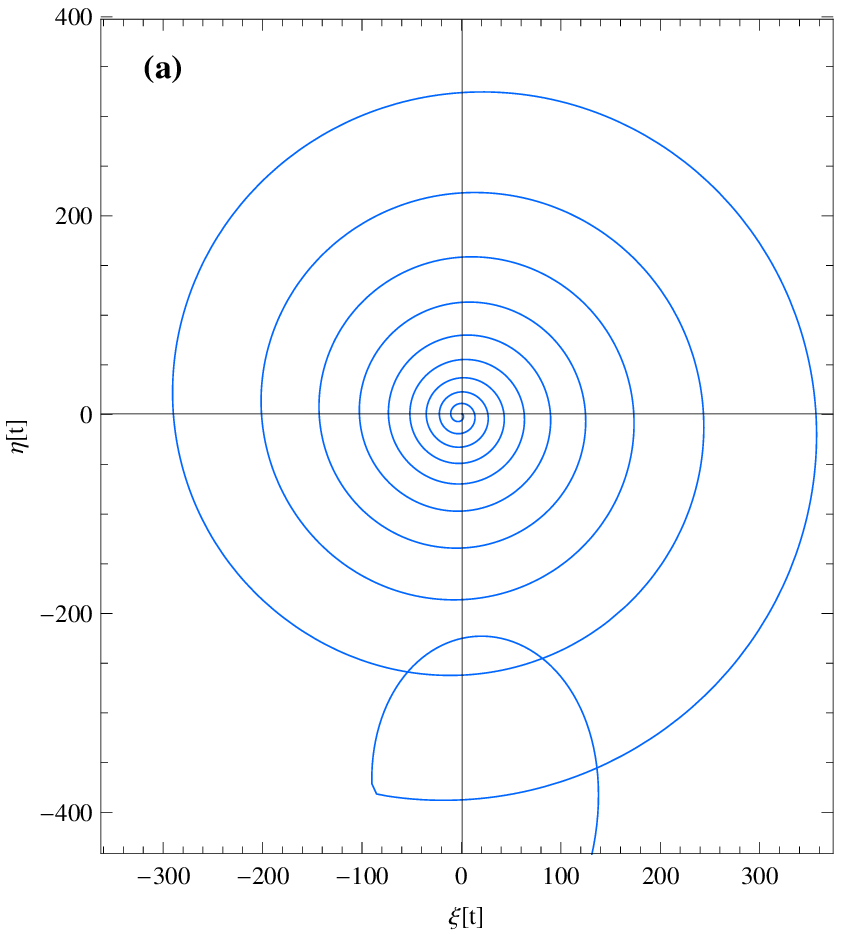}\\ \plotone{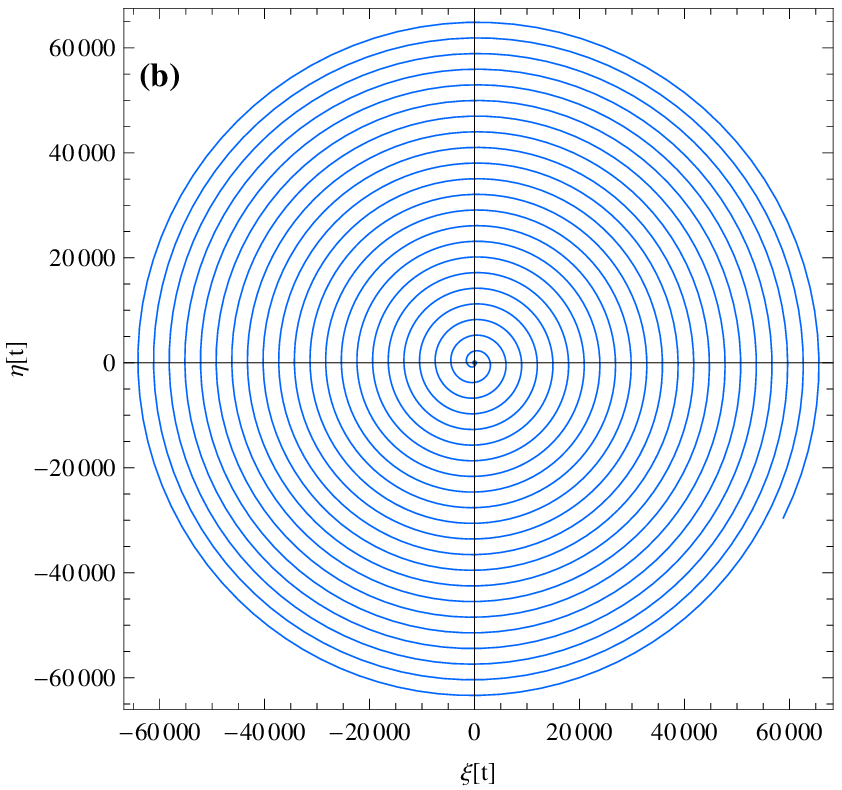}
 \caption{Trajectory of non-collinear points \label{fig:fig13}}
\end{figure}
Numerically the coordinates of triangular equilibrium points are $L_4f=(0.615824,0.75845)$, $L_5f=(0.615824,-0.75845)$ which are obtained by taking the same parametric values as  earlier taken in case of collinear equilibrium points. The parametric solution (i.e. triangular point) is depicted in figure (\ref{fig:fig13}). Frame $(a)$ of this figure shows that triangular point exists regularly for $0\leq t\leq 60$ and then a sudden change in trajectory has been seen for $60\leq t\leq 80$. Finally, solution maintains its regular path for $t>80$. Whereas frame $(b)$ of this figure shows the  nature of the path of infinitesimal body for the time interval $0\leq t\leq 200.$ It is also found that the infinitesimal mass escapes out side of possible region after $t>424229.$   
\section{Poincar\'{e} surfaces of section and Lyapunov  characteristic exponents}
\label{sec:lce}
The Poincar\'{e} surfaces of section (PSS) is an important technique to describe the stability region of the system. In order to determine the orbital element of the infinitesimal body at any instant, it is necessary to know its position ($\xi,\eta$) and velocity ($\dot \xi,\dot \eta$) correspond to a four dimensional phase space. In this paper, we have determined PSS in the $\xi\dot \xi$-plane using Event Locator Method of Mathematica$^{\textregistered}$\citep{wolfram2003mathematica}. 
The magnitude of the velocity component $\dot \eta$ is computed with the help of Jacobi integral equation (\ref{eq:ee2}) however the Jacobi constant $C$ initially computed at $t=0$. Now, we have plotted the graph of $\xi$ and $\dot \xi$ against each other when the trajectory intersects the plane in the direction of $\dot \eta>0$. In other words, a smooth well defined island in the PSS indicates that the trajectory is likely to be regular whereas the fuzzy distribution of intersection points represents chaotic trajectory. Again, if the curve shrink to a point that means it has a periodic orbit. Further, we have obtained PSS at the values of Jacobi constant $C$ for a certain range of values of $\xi$ and $\dot \xi$ whereas each orbit is determined with initial conditions:
\begin{eqnarray}
 &&\xi=\xi_0,\quad \eta=0,\quad \dot \xi=0,\nonumber\\&&
\dot \eta=\sqrt{b_3+b_4+\xi_0^2-\dot \xi_0^2-C},
\end{eqnarray}where
\begin{eqnarray}
&&b_3=\frac{2\mu_E}{(\xi_0+0.012)}+\frac{2\mu_M}{(\xi_0-0.987)}+\frac{2\mu_S(1-\beta)}{(\xi_0-389.172)} \nonumber\\&& \text{and} \nonumber\\&&
 b_4=\frac{2\beta \mu_S (1+sw)}{c}\left\{ \frac{1}{2(\xi_0-389.172)^2} \right.\nonumber \\&& \left.+arctan\left(\frac{\eta}{\xi_0-389.172}\right) \right\}.\nonumber
\end{eqnarray}
Since, in the proposed model key quantities are the values of $C, \beta$ and $sw$. Therefore, we have plotted the graphs at these parameters.
\begin{figure}[h]
 \plotone{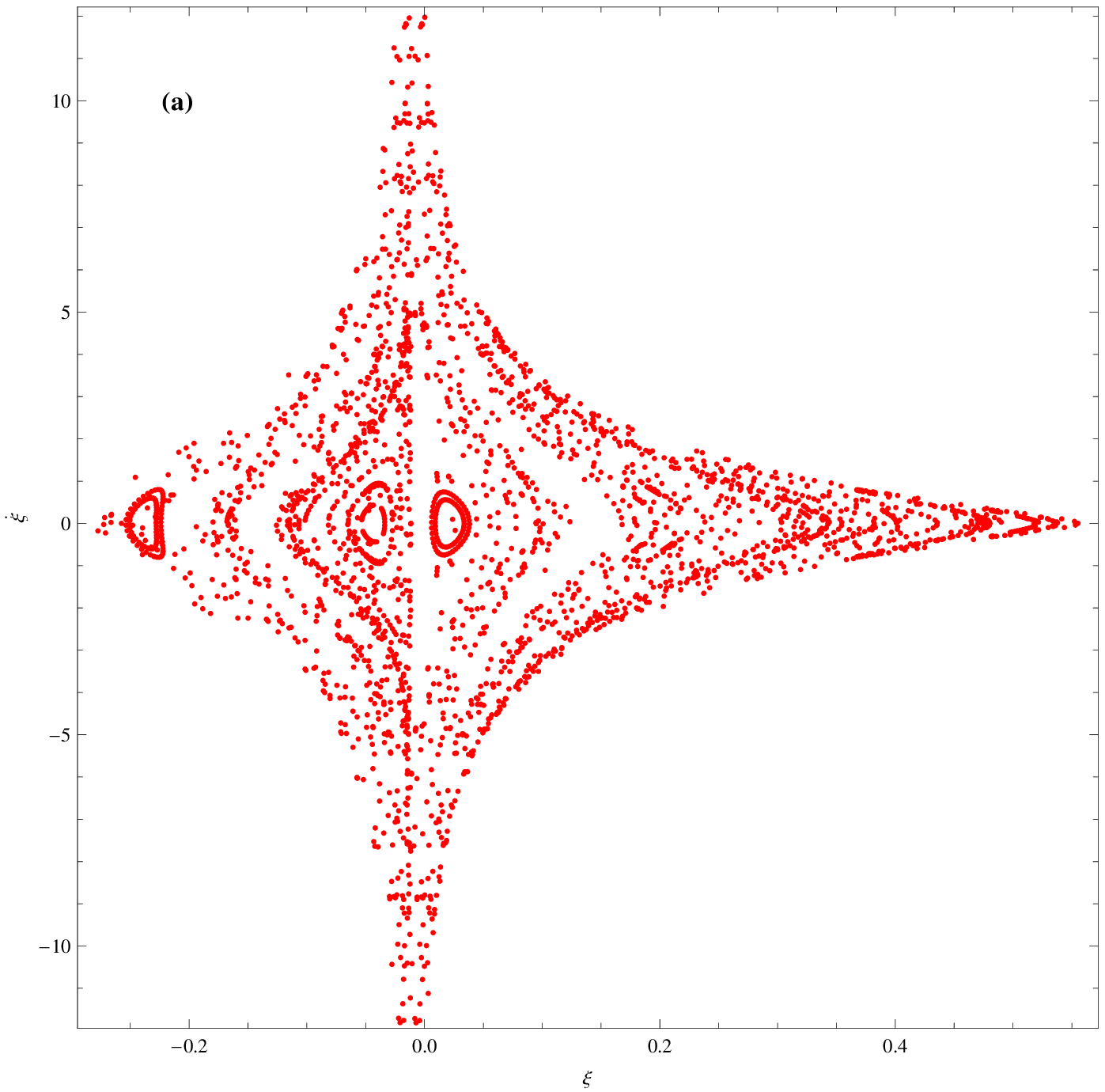}\\\plotone{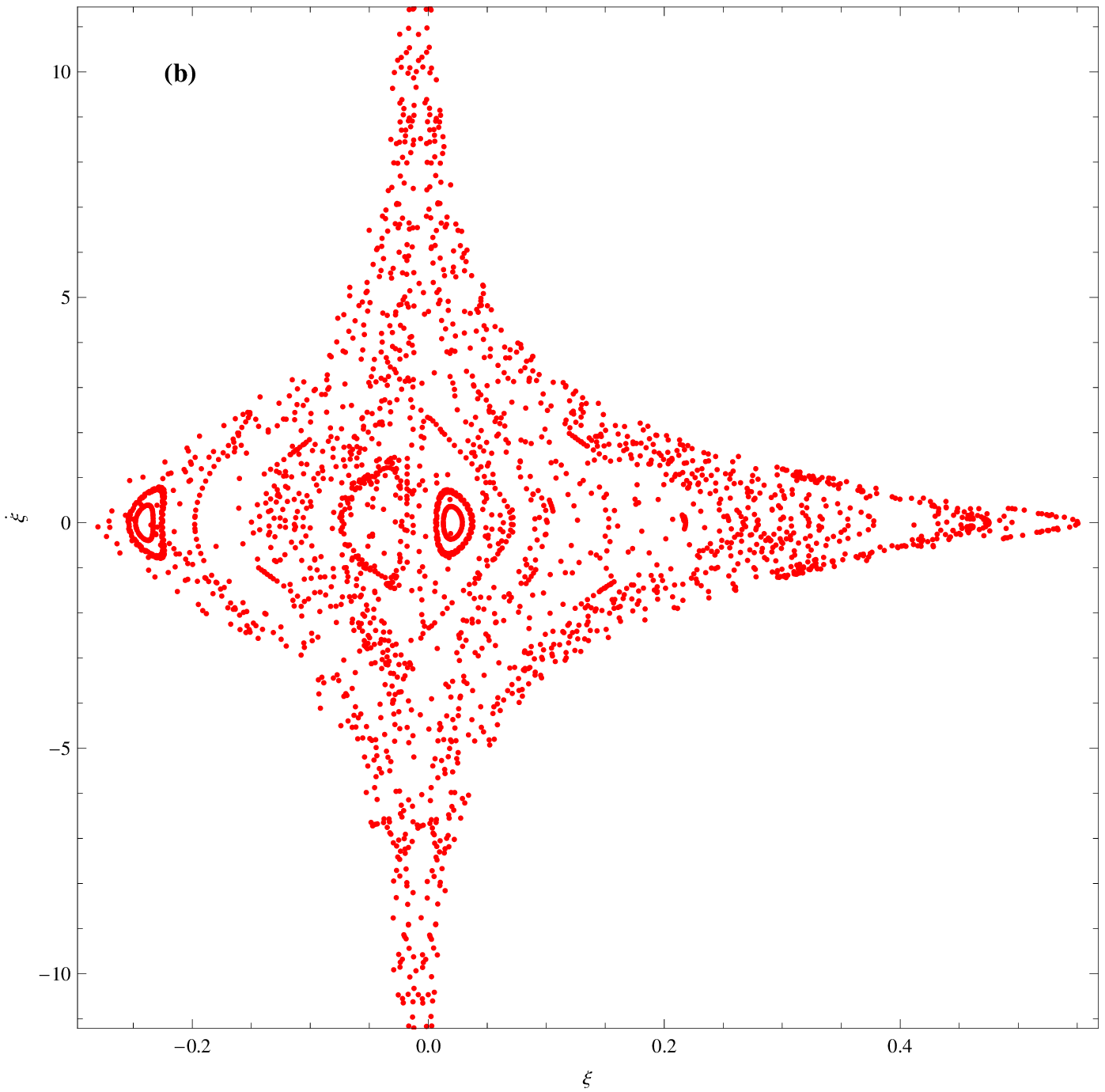}
 \caption{Poincare surfaces of section. \label{fig:fig14}}
\end{figure}
Figure (\ref{fig:fig14}) shows PSS for the restricted four body problem at the minimum values of Jacobi constant $C=1696.48$ which is initially determined at  $\xi=0.5, \eta=0,\dot \xi=0,\dot \eta=0$. In this figure, we observe that the nature of PSS and the trajectory in $\xi \dot \xi$-plane. From frame (a), we have seen regular island with the radiation pressure and the solar wind drag but frame (b) shows the regular island without any force. Here, it is clear that the regular island expands gradually due to the effect of radiation pressure and solar wind drag. Again, the island at $\xi=0.025$ shows that the trajectory is regular i.e. region in the neighborhood of $\xi=0.025$ is stable. On the other hand, in frame (b) the island at $0.025$ shows that the trajectory is regular and this region shrinks towards center. Hence from frames (a) and (b), we  conclude that the radiation pressure and solar wind drag have significantly affect on the stability region of the trajectories.

\begin{figure}[h]
\plotone{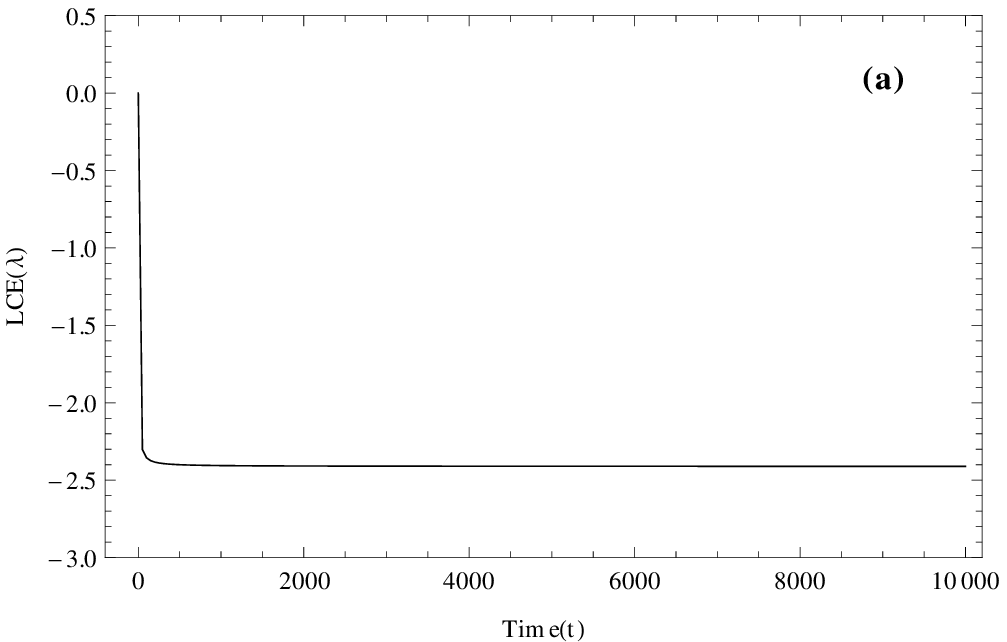}\\ \plotone{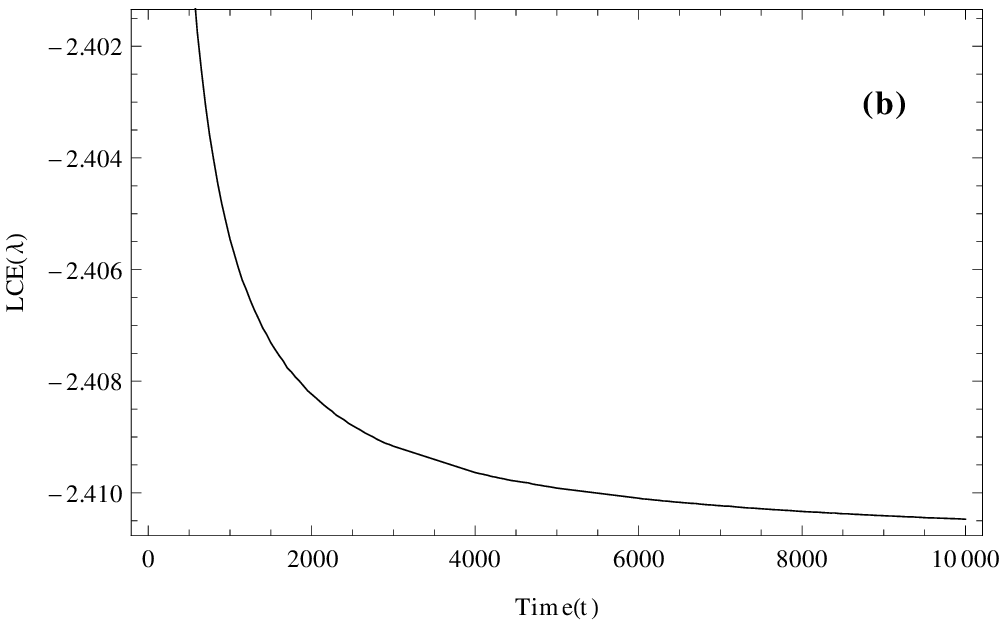}
 \caption{ Lyapunov characteristic exponent: (a) for $0\leq t\leq10000$ (b) Zoom of (a).\label{fig:fig15}}
\end{figure}
In order to know the behavior of nearby trajectory emanating  from the neighborhood of an equilibrium point we compute the LCEs.  With the help of method and algorithm described in \cite{Skokos2010LNP...790...63S}, we have numerically computed first order LCEs and plotted the graphs  $t$ Vs LCE($\lambda$), upto maximum evolution of time $0 \le t \le 10000$  when $\mu_E=0.987, \mu_M=0.012, \mu_S=328900.48, \beta=0.00002, sw=0.248, c=1.8\times 10^6$ (figure \ref{fig:fig15}). This figure shows that the proposed model is slightly affected in presence of the radiation pressure and the solar wind drag. Also, it is clear that $\lambda$  decreases slowly with time. Furthermore, $\lambda$ is negative for $0\leq t\leq 10000$ which shows that orbit is regular in this interval. Therefore, we observed that the values of LCE depends on initial deviation vector and renormalized time step. As we have used default method of Mathematica$^{\textregistered}$ software package for the numerical solutions of the differential equations.  The obtained result depends on precision and accuracy goals, i.e. in present problem, if these goals were chosen different from $3$ consequently there was overflow in the result which leads to the false estimation of the LCE.

\section{Discussion and conclusion}
\label{sec:con}
We have studied the restricted four body problem by assuming the effect of radiation pressure and solar wind drag. The boundaries of allowed regions for the motions of the infinitesimal mass are determined using zero velocity surfaces at different values of radiation pressure and solar wind drag. It is found that allowed possible regions of the motions decrease with the an increase in the value of Jacobi constant $C$. We have observed that in presence of drag forces, Jacobi constant depends on the time \citep{Murray1994Icar..112..465M,Liou1995Icar..116..186L}.  The range of radiation pressure and solar wind drag are determined and found that motion is possible for the values lie  in the interval $0<\beta<1$  and $-1<sw<1$ respectively. The curvature of the ZVC is obtained and noticed that when the curves bend to each other then connection is open and the body moves from one allowed side to other side however when the curve is not bended then curve goes outside and the body does not move one allowed side to other side. 

We have obtained the particular solutions, which are the extreme values of Jacobi function.  It is found that  three collinear points for the values of $0<\beta<0.1$ are real whereas only one real point exists  when  $0.125<\beta< 1$.  We have determined the co-ordinates of two non-collinear points  which depend on radiation pressure and the solar wind drag. With the help of  PSS, it is observed that  the stability region get expanded in presence of radiation pressure and solar wind drag and at the point $\xi=0.025$ orbits are stable.  Further, we estimated the LCE$(\lambda)$ for the maximal time of evolution $0\leq t \leq 10000$ and found that it is negative which shows that orbit of the infinitesimal body is regular. Since it is difficult to obtained an exact expression of pseudo potential function in presence of term due to dissipative force, therefore further work is needed in this regard. This work may be applicable to study the motion of test particle in coupled restricted three body problem with drag forces.

\acknowledgements{We are thankful to IUCAA, Pune for partially financial support to visit library and to use computing facility. We are also thankful to Prof. Ishwar, B.R.A. Bihar University, Muzaffarpur (India) and Mr. Ram Kishor, ISM, Dhanbad (India) for their  valuable suggestions during the preparation of the manuscript. We are grateful to anonymous referee for his constructive comment regarding improvement \citep{glasman2010science} of manuscript.}
\bibliographystyle{spbasic} 

\end{document}